\documentclass[12pt]{article}
\usepackage{cite}
\usepackage{graphicx}
\newcommand{\be}{\begin{equation}}
\newcommand{\ba}{\begin{array}}
\newcommand{\bd}{\begin{displaymath}}
\newcommand{\ee}{\end{equation}}
\newcommand{\ea}{\end{array}}
\newcommand{\ed}{\end{displaymath}}

\def\lsim{\mathrel{\rlap{\lower4pt\hbox{\hskip1pt$\sim$}}
    \raise1pt\hbox{$<$}}}      %less than or approx. symbol
\def\gsim{\mathrel{\rlap{\lower4pt\hbox{\hskip1pt$\sim$}}
    \raise1pt\hbox{$>$}}}      %greater than or approx. symbol

\def\frac#1#2{{#1\over #2}}

\def\ds {\displaystyle}

\def\xi2{$\chi^2_{d.o.f}$}
\def\x2{$\chi^2$}

\def\P {\cal P}

\def\g2{ GeV$^2$}
\def\G2{{\rm (GeV}^2)}

\def\ie{\hbox{\it i.e. }}
\def\etc{\hbox{\it etc... }}
\def\eg{\hbox{\it e.g. }}

\def\etal{\hbox{\it et al. }}

\begin{document}

%\begin{titlepage}
~

\bigskip
\bigskip

\begin{center}
{\Large \bf Pomeron effective intercept,\\logarithmic derivatives of
$F_{2}(x,Q^{2})$ \\in DIS and Regge models.}
\bigskip
\bigskip
\end{center}

\begin{center}

{\large {
P. Desgrolard $^{a,}$\footnote{E-mail:
desgrolard@ipnl.in2p3.fr},
A. Lengyel $^{b,}$\footnote{E-mail:
sasha@len.uzhgorod.ua},
E. Martynov $^{c,}$\footnote{E-mail:
martynov@bitp.kiev.ua} }}

\end{center}

\bigskip
\bigskip

\noindent
$^a$ Institut de Physique Nucl\'eaire de Lyon, IN2P3-CNRS et
Universit\'e Claude Bernard, 43 boulevard du 11 novembre 1918, F-69622
Villeurbanne Cedex, France

\noindent $^b$ Institute of Electron Physics, National Academy of
Sciences of Ukraine, 88015 Uzhgorod-015, Universitetska 21, Ukraine

\noindent $^c$ Bogoliubov Institute for Theoretical Physics, National
Academy of Sciences of Ukraine, 03143 Kiev-143, Metrologicheskaja 14b,
Ukraine

\bigskip
\bigskip

\begin{center}
\begin{minipage}[t]{13.0cm}
\noindent {\bf Abstract} The drastic rise of the proton structure
function $ F_2(x,Q^2)$ when the Bj\"orken variable $x$ decreases, seen
at HERA for a large span of $Q^2$, may be damped when $x\to 0$ and
$Q^2$ increases beyond $\sim $ several hundreds \g2. This phenomenon
observed in the Regge type models is discussed in terms of the
effective Pomeron intercept and of the derivative $B_x=\partial{\ell n
F_2(x,Q^2)}/\partial{\ell n(1/x)}$. The method of the overlapping bins
is used to extract the derivatives $B_x$ and $B_Q=\partial{\ell n
F_2(x,Q^2)}/\partial{Q^2}$ from the data on $F_2$ for $6.0\cdot
10^{-5}\le x\le 0.61$ and $ 1.2 \le Q^2 $ (\g2) $\le 5000$. It is shown
that the extracted derivatives are well described by recent Regge
models with the Pomeron intercept equal one.
\end{minipage}
\end{center}

\bigskip

%%%%%%%%%%%%%%%%%%%%%%%%%%%%%%%%%%%%%%%%%%%%%%%%%%%%%%%%%%%
%%%%%%%%%%%%%%%%%%%%%%%%%%%%%%%%%%%%%%%%%%%%%%%%%%%%%%%%%%%
%%%%%%%%%%%%%%%%%%%%%%%%%%%%%%%%%%%%%%%%%%%%%%%%%%%%%%%%%%%
\section{\Large{Introduction}}
The so-called "HERA effect" discovered quite long ago by the
experimentalists in deep inelastic scattering (DIS) and anticipated by
theoreticians has arisen a great interest. It concerns the strong rise of
the proton structure functions (SF) $F_2(x,Q^2)$ when the Bj\"orken
variable $x$ decreases, for the experimentally investigated $Q^2$,
negative values for the 4-momentum transfer.

This phenomenon is described usually within assumption about power-like
behavior of SF at small x
\begin{equation}\label{power}
F_{2}\ \propto x^{-\lambda} \qquad \mbox{where} \quad \lambda>0.
\end{equation}
It is assumed as a rule that the power $\lambda$ is related with the
intercept of the Reggeon contribution dominating at $x\to 0$, namely with
the Pomeron intercept
\begin{equation}\label{lambda}
\lambda=\alpha_{\P}(0)-1.
\end{equation}
One can find a justification of this treatment in pQCD where the well
known BFKL Pomeron in the LL approximation has
\begin{equation}\label{lambda_BFKL}
\lambda_{BFKL}=3N_{c}(\alpha_{s}/\pi)\ell n2\approx 0.4.
\end{equation}
However it is necessary to have in mind two circumstances. Firstly, the
next-to-leading correction to $\lambda_{BFKL}$ is negative and quite
large \cite{FadLip}. Generally, the BFKL Pomeron is only a perturbative
approximation to the true Pomeron of which exact properties as well as
properties of the whole perturbative series are unknown. Secondly. a
power-like $x$-dependence of $F_{2}$ with $\lambda =const>0$ leads to
the violation of unitarity providing the Froissart-Martin \cite{F-M}
bound for total cross-section is valid for $\gamma^{*} p$ interaction
at least at $x\to 0$. These facts make a ground for a widely accepted
opinion that the BFKL Pomeron is to be transformed into a
$j$-singularity with intercept exactly equaled to one when all
corrections are taken into account~\footnote{Of course, one should not
forget that Froissart-Martin bound is not proved yet for $\gamma p$
scattering amplitude.}. If that is true a power-like behavior of SF at
small-$x$ will be changed for a logarithmic-like one (similarly to
hadron amplitudes, when a procedure of unitarization or eikonalization
is applied). Evidently, such an evolution should lead to a damping of
the fast growth of SF at $x\to 0$. We think it is an interesting and
important task to investigate the available data to see if they exhibit
such a tendency which can be named as damping of the HERA effect.

The detailed analysis \cite{lamb(Q2)} of experimental data shows an
evident dependence of $\lambda$ on $Q^{2}$. This property of $\lambda$
contradicts  relation (\ref{lambda}) provided the Pomeron is the universal
object as it is observed in pure hadronic reactions \cite{CEKLT,CEGKKLNT}.
The Pomeron trajectory and hence its intercept must be independent of the
external interacting particles, the true universal Pomeron intercept
$\alpha_{\P}(0)\leq 1$ does not depend on $x$ or $Q^{2}$. It means that a
simplified power-like parameterization (\ref{power}) (with a constant
$\lambda >0$) of the forward $\gamma^{*}p$ amplitude or structure function
should be considered only as an effective one and can be applied only in
narrow intervals of $x$ and $Q^{2}$.

Following the suggestion of \cite{NPW} the logarithmic $x-$derivative
of the structure function (let note it as $B_{x}(x,Q^{2})$ and name it
as $x$-slope for brevity)
\begin{equation}\label{x-slope}
B_{x}(x,Q^{2})=\frac{\partial \ell nF_{2}(x,Q^{2})}{\partial \ell
n(1/x))}.
\end{equation}
was recently extracted \cite{H1Rep,BH1-01} from the data on
$F_{2}(x,Q^{2})$. However the wrong relation between $x-$slope and "{\it
Pomeron effective intercept}" was used in the report \cite{H1Rep}.
Starting from the structure function written in the form
\begin{equation}\label{SF}
F_{2}(x,Q^{2})=G(Q^{2})\left ( \frac{1}{x}\right)^{\lambda(x,Q^{2})}
\end{equation}
one can define the Pomeron effective intercept as follows
\begin{equation}\label{effint}
\alpha_{\P}(x,Q^{2})\equiv 1+\lambda(x,Q^{2}).
\end{equation}
"Experimental" data on effective intercept have been extracted in \cite{H1Rep}
from the data on structure function $F_{2}(x,Q^{2})$ using the
relation
\begin{equation}\label{x_der-wrong}
\lambda(x,Q^{2})=\frac{\partial \ell nF_{2}(x,Q^{2})}{\partial \ell
n(1/x)}\quad (=\ B_{x}(x,Q^{2})).
\end{equation}
The correct equation relating $x$-slope, $B_{x}$, and
$\lambda(x,Q^{2})$ must be read however as follows
\begin{equation}\label{x_der-corr}
\frac{\partial \ell nF_{2}(x,Q^{2})}{\partial \ell
n(1/x)}=\lambda(x,Q^{2})+\frac{\partial \lambda(x,Q^{2})}{\partial \ell
n(1/x)}.
\end{equation}
The relations (\ref{x_der-wrong}) and (\ref{x_der-corr}) strictly coincide
only if $\lambda $ does not depend on $x$.  We must note
that one of the conclusions made in \cite{H1Rep,BH1-01} is
that experimental data support an independence of $B_{x}(x^{2},Q^{2})$ at
$x<0.01$. Nevertheless, in our opinion one need to be more careful.

Supposing that $B_{x}(x,Q^{2})$ does not depend on $x$ at small $x$ we
have the following equation for $\lambda(x,Q^{2})$
\begin{equation}\label{lambda-eq}
\lambda '(x,Q^{2})+\lambda(x,Q^{2})=\lambda_{c}(Q^{2})
\end{equation}
where the "prime" denotes a derivative with respect to $\ell n(1/x)$.
There are two possible solutions of this equation.

-1) $\lambda'\neq 0$. In this case
$$
\lambda(x,Q^{2})=\lambda_{c}(Q^{2})(1-\ell n(1/x))
$$
and $\lambda(x,Q^{2})\to -\infty$ at $x\to 0\ $ (if $\lambda_{c}(Q^{2})>0$).
Such a behavior seems very unnatural.

-2) $\lambda'=0$. In this case $$\lambda(x,Q^{2})=\lambda_{c}(Q^{2}),$$
\ie it does not depend on $x$ and hence $\sigma_{tot}^{\gamma
p}(W)\propto W^{2\lambda_{c}(0)}$ at $W\to \infty$. If we believe in
the Pomeron universality we should rather reject than accept this
solution as valid one for \underline {arbitrary} small $x$.

Thus the conclusion about constant $B_{x}(x,Q^{2})$ is possible due to
large enough errors and dispersion of the obtained data on $B_{x}$. In
fact, it is more likely that $x$-slope depends weakly on $x$, \ie
$$
\left |\frac{\partial \lambda(x,Q^{2})}{\partial \ell n(1/x)}\right |\ll
\lambda(x,Q^{2}).
$$
However it is difficult without any model assumptions to determine
quantitatively how much the derivative of $\lambda$ is small. As we show
in the Section \ref{Comparison}, the data on $x$-slope are well described
in the models with $\alpha_{P}(x,Q^{2})$ and hence with $\lambda(x,Q^{2})$
depending on $x$.

At the same time it seems possible in principle to derive some
conclusions about effective intercept from the behavior of
$B_{x}(x,Q^{2})$ at small $x$ and high $Q^{2}$. In particular at fixed
(even high) $Q^{2}$ and $x\to 0$ unitarized Regge models predict a
decreasing of $x$-slope and hence of $\lambda(x,Q^{2})$ \cite{TroTyu}.
We discuss this subject in more details in the Section \ref{Models}.

The data on $x$-slope have been given by H1 Collaboration \cite{H1Rep} only
in the form of figures and cannot be used yet for quantitative
comparison with models\footnote{When the present paper was practically
completed, data on $B_{x}$ extracted by H1 Collaboration have been
published \cite{BH1-01}. We compare these data with our results in the
Section \ref{Data}} predictions. Therefore we have performed our own
procedure of extracting the local slopes $B_{x}(x,Q^{2})$ for fixed
$Q^{2}$ as well as for fixed $x$ based on the method of so-called
"overlapping bins". A brief description of the method and results are
presented in the Section \ref{Data}.

In addition to the $x$-slope, we consider another logarithmic
derivative of the SF, noted $B_{Q}(x,Q^2)$, named $Q$-slope and defined by
\begin{equation}\label{B-Q}
B_{Q}(x,Q^{2})\equiv \frac{\partial F_{2}(x,Q^{2})}{\partial \ell nQ^{2}}.
\end{equation}
The merit of this slope is mainly due to its relation with the gluon
density at $x\ll 1$. The so-called Caldwell plot \cite{Caldwell} was
the first presentation of the ZEUS data on $B_{Q}$ as a useful tool
to study scaling violation of SF and interrelation between soft
and hard physics in DIS. However as repeatedly noted
\cite{GotLevMaor,DLM,DJLP,DLM-bx,KMP}, the data shown in the original
Caldwell plot are derived from $F_{2}$ when averaging over
rather large intervals in $Q^{2}$ and for correlated values of $x$ and
$<Q^{2}>$. Taking into account the importance of this quantity, it would
be useful to have in a wide region of $x$ and $Q^{2}$ the local rather
than averaged values of $B_{Q}(x,Q^{2})$. Using the same method of
overlapping bins we extract the local $Q$-slopes not only from the
H1-data (it was made in \cite{H1-00}) but also from the
ZEUS-data and compare  with the predictions of our models. The results are
presented in the Section~\ref{Comparison}.

%%%%%%%%%%%%%%%%%%%%%%%%%
%%%%%%%%%%%%%%%%%%%%%%%%%
%%%%%%%%%%%%%%%%%%%%%%%%%
\section{Regge Models for proton structure function}\label{Models}
There are many phenomenological models \eg
\cite{TroTyu,DLM,DL,ALLM,Be-95,CKMT,PetPro,CFSK,DM,CS} developed within
the Regge approach for $\gamma^{*}p$ scattering and structure
functions. Some of them \cite{TroTyu,DL,ALLM,CKMT,PetPro} deal with a
Pomeron having intercept $\alpha_{P}(0)>1$ and can be considered as
input for further unitarization \cite{TroTyu,PetPro,CFSK}. Other ones
are constructed from the beginning as models do not violating the
Froissart-Martin bound for total cross-sections. Let remind we support
the Pomeron universality meaning that, if the Pomeron
contribution in pure hadronic amplitudes satisfies unitarity
restrictions, it should have the same feature in $\gamma$-hadron
scattering.

We consider here two types of unitarized Pomeron, leading
at $s\to \infty$ to $\sigma_{tot}(s)\propto \ell ns$ \cite{DLM,DM} and to
$\sigma_{tot}(s)\propto \ell n^{2}(s)$ \cite{CS}.

%%%%%%%%%%%%%%%%%%%%%%%%%%%%%%%%%%%%%%%%%%%%%%%%%%%5
%%%%%%%%%%%%%%%%%%%%%%%%%%%%%%%%%%%%%%%%%%%%%%%%%%%%
\subsection{Soft Dipole Pomeron (SDP) model}\label{SDP}
Defining the Soft Dipole Pomeron model for DIS \cite{DLM,DM}, we start
from the expression relating the transverse cross-section of $\gamma^*p$
interaction with the proton structure function $F_2$ and the optical
theorem for forward scattering amplitude.
\begin{equation}
 \sigma_{T}^{\gamma^*p}(W^2,Q^2)=8\pi\Im m A(W^2,Q^2;t=0)=
\frac{4\pi^2\alpha}{Q^2(1-x)}(1+4m_p^2x^2/Q^2) F_2(x,Q^2)\ ;
\end{equation}
the longitudinal contribution to the total cross-section,
$\sigma_L^{\gamma^*p}=0$ is assumed.

The forward scattering at $W$ far from the $s$-channel threshold
$W_{th}=m_p$ is dominated by the Pomeron and the $f$-Reggeon
\begin{equation}\label{ampl}
 A(W^2,t=0;Q^2)=P(W^2,Q^2)+f(W^2,Q^2),
\end{equation}
with the Reggeon contribution
\begin{equation}\label{f-term}
 f(W^2,Q^2)=iG_f(Q^2)(-iW^2/m_p^2)^{\alpha_f(0)-1}(1-x)^{B_f(Q^2)}\ ,
\end{equation}
where
\begin{eqnarray}
G_f(Q^2)=\frac{C_f}{\left(1+Q^2/Q_{f}^2
\right)^{D_f(Q^2)}}\ ,\label{G-f} \\
D_f(Q^2)=d_{f\infty}+\frac{d_{f0}-d_{f\infty}}{1+Q^2/
Q_{fd}^2}\ ,\label{D-qf}
\end{eqnarray}
\begin{equation}\label{B-xf}
B_f(Q^2)=b_{f\infty}+\frac{b_{f0}-b_{f\infty}}{1+Q^2/Q^2_{fb}}\ .
\end{equation}
For the Pomeron contribution, we take a two-component form where the two
Pomerons have an intercept equaled to one: one of both being a simple
$j$-pole while the other one is a double $j$-pole.

\begin{equation}\label{pom-term}
P(W^2,Q^2)=P_1+P_2\ ,
\end{equation}
with
\begin{eqnarray}
P_1&=&iG_1(Q^2)\ell n(-iW^{2}/m_{p}^{2})(1-x)^{B_1(Q^2)},\label{p-1} \\
P_2&=&iG_2(Q^2) (1-x)^{B_2(Q^2)},\label{p-2}
\end{eqnarray}
where
\begin{eqnarray}
G_i(Q^2)&=&\frac{C_i}{\left(1+Q^2/Q_{i}^2
\right)^{D_i(Q^2)}}\ , \qquad i=1,2,\label{G-ip} \\
\qquad D_i(Q^2)&=&d_{i\infty}+\frac{d_{i0}-d_{i\infty}}{1+Q^2/ Q_{id}^2}\ ,
\qquad i=1,2,\label{D-qp}
\end{eqnarray}
\begin{equation}\label{B-xp}
B_i(Q^2)=b_{i\infty}+\frac{b_{i0}-b_{i\infty}}{1+Q^2/Q^2_{ib}}\ , \qquad
i=1,2.
\end{equation}
We would like to comment the above expressions, especially the powers
$D_{i}$ and $B_{i}$ varying smoothly between constants when $Q^{2}$ goes
from $0$ to $\infty$. In spite of an apparently cumbersome form they are a
direct generalization of the exponents $d$ and $b$ appearing in each term
of the simplest parametrization of the $\gamma^*p$-amplitude
$$
G(Q^2)=\frac{C}{(1+Q^2/Q_0^2)^d}\qquad \mbox{and}\qquad (1-x)^b\ .
$$
Indeed, a fit to experimental data shows unambiguously that the parameters
$d$ and $b$ should depend on $Q^2$.

Details of the fit to the data and values of fitted parameters can
be found in \cite{DLM,DM}. Here, we only note that in the region
$W>3$ GeV$^{2}$,
$0\leq Q^{2}\leq 3000$ GeV$^{2}$ and $x\leq 0.75$ we obtained
$\chi^{2}/dof=1.073$, where and in what follows "$dof$" means "degree of
freedom" = number of experimental points $-$ number of free parameters.

We concentrate on asymptotic behavior of $x$-slope, $B_{x}(x,Q^{2})$. At
very small $x$ term $P_{1}$ dominates in (\ref{ampl}). Hence at $x\ll 1$
\begin{equation}\label{B-xSDP}
\begin{array}{lll}
B_{x}(x,Q^{2})=\ds \frac{\partial{\ell n F_2(x,Q^2)}}{\partial{\ell
n(1/x)}}&\approx &\ds \frac{\partial \ell n(K(x,Q^{2})P_{1})}{\partial
\ell n(1/x)}\approx \frac{\partial \ell n\left (\ell
n(W^{2}/m_{p}^{2})(1-x)^{B_{1}(Q^{2})}\right )}
{\partial \ell n(1/x)}\\
&\approx &\ds \frac{1}{\ell n(W^{2}/m_{p}^{2})}\approx \frac{1}{\ell
n(Q^{2}/xm_{p}^{2})}\ ,
\end{array}
\end{equation}
where
$$
K(x,Q^{2})=\frac{2\pi (1+4m_{p}^{2}x^{2}/Q^{2})}{\alpha Q^{2}(1-x)}\ .
$$
Thus in both cases, at fixed $Q^{2}$ and $x\to 0$ as well as at fixed
$x$ and $Q^{2}/m_{p}^{2}\gg 1$ (but provided that $P_{1}$ term dominates),
in the SDP model $B_{x}(x,Q^{2})\ll 1$. Furthermore, as we show in
Section \ref{Comparison}, the available data on $x$-slope and its growth
with $Q^{2}$ and $1/x$ are very well described within the model. The
decreasing of $B_{x}$ is predicted far out of HERA kinematic
region.

%%%%%%%%%%%%%%%%%%%%%%%%%%%%%%%%%%%%%%%%%%%%%%%%%%%%%%%%
%%%%%%%%%%%%%%%%%%%%%%%%%%%%%%%%%%%%%%%%%%%%%%%%%%%%%%%%
\subsection{Generalized Logarithmic Pomeron (GLP) model}\label{GLP}
In \cite{DM} we have constructed a model incorporating a slow rise
of $\sigma_{tot}^{\gamma p}(W^{2})$ and simultaneously a fast rise of
$F_{2}(x,Q^2)$ at large $Q^{2}$ and small $x$. The model is in a sense
intermediate between the above Soft Dipole Pomeron model and the
two-Pomeron model \cite{DL} of Donnachie and Landshoff
.
The model is defined as follows.
\begin{equation}\label{LogG}
F_{2}(x,Q^{2})=F_{0}+F_{s}+F_{f},
\end{equation}
\begin{equation}\label{LogG-0}
F_{0}=C_{0}\frac{Q^{2}}{\left(1+Q^{2}/Q^{2}_{0}\right)^{d_{0}}}
(1-x)^{B_{0}(Q^{2})},
\end{equation}
\begin{equation}\label{LogG-s}
F_{s}=C_{s}\frac{Q^{2}}{(1+Q^{2}/Q^{2}_{s })^{d_{s}}}L(W^{2},Q^{2})
(1-x)^{B_{s}(Q^{2})},
\end{equation}
where
\begin{equation}\label{L(Q,W)}
L(W^2,Q^{2})=\ell n \left[1+\frac{a}{(1+Q^{2}/Q^{2}_{s\ell})^{d_{s\ell}}}
\left(\frac{Q^{2}}{xm_{p}^{2}}\right)^{\epsilon}\right]\ ,
\end{equation}
\begin{equation}\label{LogG-f}
F_{f}=C_{f}\frac{Q^{2}}{(1+Q^{2}/Q^{2}_{f})^{d_{f}}}
\left(\frac{Q^{2}}{xm_{p}^{2}}\right)^{\alpha_{f}(0)-1}(1-x)^{B_{f}(Q^{2})}
\end{equation}
and
\begin{equation}\label{LogG-B}
B_{i}(Q^{2})=b_{i\infty}+\frac{b_{i0}-b_{i\infty}}{1+Q^{2}/Q^{2}_{ib}},
\qquad i=0, s, f.
\end{equation}
For $\gamma p$ total cross-section the model gives
\begin{equation}\label{glpsigt}
       \sigma_{tot}^{\gamma p}(W^2)= 4\pi^2\alpha\left[
           C_0+C_s L(W^2,0)+C_f\left({W^2\over m_p^2}-1\right)^
           {\alpha_f(0) -1}\right]\ ,
\end{equation}
        with
$$
      L(W^2,0)\ =\ \ell n\left(1+a\left({W^2\over m_p^2}-1\right)
         ^{\ \epsilon}\right)\ .
$$
A few comments on the above model are needed.

The new logarithmic factor in (\ref{LogG-s}) can be rewritten in the form
$$
L(W^2,Q^{2})=\ell
n\left[1+\frac{a}{(1+Q^{2}/Q^{2}_{s\ell})^{d_{s\ell}}}\left(
\frac{W^{2}+Q^{2}}{m_p^{2}}-1\right)^{\epsilon}\right].
$$
Consequently, at $Q^{2}=0$, we have $L(W^{2},0)\approx \epsilon\ell
n(W^{2}/m_p^{2})$ at $W^{2}/m_p^{2}\gg 1$. Thus, $\sigma_{tot}^{\gamma
p}(W^{2})\propto \ell n W^{2}$ at $W^{2}\gg m_p^{2}$. A similar behaviour
can be seen at moderate $Q^{2}$ when the denominator is $\sim 1$. However
at not very large $W^{2}/m_p^{2}$ or at sufficiently high $Q^{2}$ the
argument of logarithm is close to 1, and then
$$
L(W^2,Q^{2})\approx \frac{a}{(1+Q^{2}/Q^{2}_{s\ell})^{d_{s\ell}}}\left(
\frac{W^{2}+Q^{2}}{m_p^{2}}-1\right)^{\epsilon}\ ,
$$
simulating a Pomeron contribution with intercept
$\alpha_{P}(0)=1+\epsilon$.

We are going to justify that, in spite of its appearance, the GLP model
cannot be treated as a model with a hard Pomeron, even when $\epsilon$
issued from the fit is not small. In fact, the power $\epsilon $ inside
the logarithm is NOT really the intercept (more exactly it is not
$\alpha_{P}(0)-1$). Intercept is defined as position of singularity of the
amplitude in $j$-plane at $t=0$. It is shown in \cite{DM} that in the
present GLP model the true leading Regge singularity is located exactly at
$j=1$: it is a double pole due to the logarithmic dependence.

Thus this model should be considered rather as a Dipole Pomeron model.
In order to distinguish it from the Soft Dipole Pomeron model
of Section \ref{SDP}, we have named this model "Generalized
Logarithmic Pomeron" (GLP) model. A fit performed \cite{DM} has given
$\chi^{2}/dof=1.064$. Parameters of the GLP model are presented in
\cite{DM}; for the power $\epsilon$ we obtained $\epsilon=0.4536\pm
0.0015$. Finally, we can say that the GLP model only mimics a contribution of
a hard Pomeron although non incorporating explicitely one. The asymptotic
behavior of $x$-slope at fixed $Q^{2}$ and $x\to 0$ in the GLP
model coincides with SDP one:
$$
B_{x}(x\to 0,Q^{2})\approx \frac{1}{\ell n(Q^{2}/xm_{p}^{2})}\to 0.
$$
In an other kinematical limit: $x\ll 1$, but fixed, and $Q^{2}
\to \infty$, the GLP model gives a behavior
\begin{equation}\label{B-xGLP}
B_{x}(x\ll 1,Q^2\to\infty)\approx \epsilon\left
(1-\Phi(Q^{2})\left(\frac{1}{x}\right)^{\epsilon}\right)\ ,
\end{equation}
where
\begin{equation}\label{Phi}
\Phi(Q^{2})=\frac{a(Q^{2}/m_{p}^{2})^{\epsilon}}
{(1+Q^{2}/Q^{2}_{sl})^{d_{sl}}}\propto \left(\frac{Q^{2}}{m_{p}^{2}}
\right)^{\epsilon-d_{sl}}\to 0\qquad \mbox{at} \qquad Q^{2}\to \infty\ ;
\end{equation}
because in accordance with the fit \cite{DM} $d_{sl}=0.7016\pm 0.0022$ and
hence ($\epsilon$ is given above) $\epsilon -d_{sl}\approx -0.25$.

%%%%%%%%%%%%%%%%%%%%%%%%%%%%%%%%%%%%%%%%%%%%%%%
%%%%%%%%%%%%%%%%%%%%%%%%%%%%%%%%%%%%%%%%%%%%%%%
\subsection{Cudell-Soyez Pomeron (CSP) model}\label{CSP}
Recently an other model obeying unitarity restrictions for Pomeron
contribution was suggested in \cite{CS} to account for HERA data (\ie low $x$
structure functions). The model has the form
\begin{equation}\label{Cud-Soy}
F_{2}(Q^{2}/2\nu,Q^{2})=a(Q^{2})\ell
n^{2}\left(\nu/\nu_{0}(Q^{2})\right)+c(Q^{2})+d(Q^{2})
(2\nu)^{\alpha_{f}(0)-1}\ ,
\end{equation}
where $\nu=Q^{2}/(xm_{p}^{2})$ and leads to
\begin{equation}\label{sigtot-CS}
\sigma_{tot}^{\gamma p}(s)\propto \ell n^{2}(s) \qquad \mbox{at} \qquad
s\to \infty\ .
\end{equation}
The explicit form of functions $a(Q^{2}), \nu_{0}(Q^{2}), c(Q^{2})$ and
$d(Q^{2})$ can be found in the original paper \cite{CS}. This model is
particularized by construction to the region of small $x$-SF. We have
refitted it to our set of data \cite{DM} restricted to $x\leq 0.07$ and
$0\leq Q^{2}\leq 3000$ GeV$^{2}$ and obtained $\chi^{2}/dof \approx
1.0$.
One can show that at fixed $Q^{2}$ and $x\to 0$ the slope
$B_{x}(x,Q^{2})$ calculated in CSP model is exactly twice as large as the
slope calculated in SDP model although they practically coincide
at the available $x<0.03$ and $Q^{2}<1000$ GeV$^{2}$ (see Section~
\ref{Comparison}).

%%%%%%%%%%%%%%%%%%%%%%%%%%%%%%%%%%%%%%%%%%%%%%%%%%%%%%%%%%%%%%%%%%%%
%%%%%%%%%%%%%%%%%%%%%%%%%%%%%%%%%%%%%%%%%%%%%%%%%%%%%%%%%%%%%%%%%%%%
%%%%%%%%%%%%%%%%%%%%%%%%%%%%%%%%%%%%%%%%%%%%%%%%%%%%%%%%%%%%%%%%%%%%
\section{Extraction of the local slope $B_{x}(x,Q^{2})$ from the SF
data}\label{Data}

As noted in the introduction, the $x$-slope is a precious tool to settle
if, either yes or no, a damping of the HERA effect does exist. Actually,
we need the sets of experimental data on $B_{x}(x,Q^{2})$ as a function of
$x$ at fixed $Q^{2}$ as well as at fixed $x$ versus sufficiently high
$Q^{2}$. Unfortunately, because the $x-$slope is not a measurable
observable it should be extracted, when possible, from the available data
on the SF.

To extract $x$-slopes with a good accuracy we adapt the so-called method
of "overlapping bins"~\cite{konlen}, originally intended for analyzing the
local nuclear slope of the first diffraction cone in $pp$ and $\bar pp$
elastic scattering. Then the method was used to determine averaged
$x$-slopes \cite{DJLP} and the local ones \cite{DLM-bx} before new, more
precise and complete, data appeared. Let briefly describe the main idea of
the method.
\smallskip

Provided that the SF has been measured for a given $Q^{2}$ at $N$
$x$-points lying in some interval $[x_{min},x_{max}]$, we adopt the
following procedure. First, we divide this interval into subintervals
or elementary "bins" (with $n_{b}$ measurements in each of them,
assumed for simplicity to be the same for all bins). Once the first bin
is chosen, the second bin is obtained from the first one by shifting
only one point of measurement (of course one could shift by any number
of points less or equal $n_{b}$, the shift of one point we choose is
the minimal one giving rise to the maximal number of overlapping bins).
The third bin is obtained from the second bin by the shift of one data
point \etc . Thus, we define $N-n_{b}+1$ overlapping bins for a given
$Q^2$. For each ($k$-th) bin, $n_{b}$ must be large enough and its
width (in $x$) small enough to allow fitting the SF with the simplest
form directly involving the $x$-slope
\begin{equation}\label{localF2}
  F_{2}(x)=A\left (\frac{1}{x}\right)^{B}\ ,\quad (\mbox{for a given
  fixed}\quad  Q^2)\ .
\end{equation}
The parameter $B$ represents the value of the $x$-slope
$B\left(<x>_{k},Q^{2}\right)$, "measured", at $Q^2$ and at the "weighted
average" $<x>_{k}$ defined in the $k$-th bin as (see \cite{br99})
\begin{equation}\label{average x}
  <x>_k=exp\bigg (-\frac{\sum\frac{\ln x_{i}}{\Delta y_{i}}}
  {\sum\frac{1}{\Delta y_{i}}}\bigg)\ ,\quad
  k\in{[1,N-n_{b}+1]}\      ,
\end{equation}
where $x_{i}$ is the value of $x$ at which the structure function $y_{i}$
is measured with the uncertainty $\Delta y_{i}$; the summations run over
all data points, $i=1,2,...,n_{b}$ of the bin. This yields the
"experimental" values of $B_{x}(x,Q^{2})$ with the corresponding standard
errors determined in the fit of (\ref{localF2}) to the data. Then the
procedure is to be repeated for all bins and ultimately for the other
$Q^{2}$'s at which the SF have been measured.

\smallskip

The next step in extracting and analyzing $B_{x}(x,Q^{2})$ is the
determination of the slopes at fixed $x$ as function of $Q^{2}$, making
use of results of the first step. As a rule, the sets of $<x>_{k}$, at
different $Q^{2}$, do not coincide. So in order to get the $x-$slope at
fixed $x$ and at different $Q^{2}$ we interpolate (or extrapolate but not
far from the $x$-interval under consideration) already extracted $B_{x}$
at the given $Q^{2}$ to the chosen $x$. This has been made simply assuming a
linear $x-$dependence of $B_{x}$.

The above method of overlapping bins is applied to the whole
available data set\footnote{The detailed description of the data used in
our analysis as well as the references for them can be found in our
previous work~\cite{DM}}. We have separated the data in three groups: ZEUS,
H1 and Fixed Target experiments which were analyzed independently
each of the others.
\smallskip
The resulting values of $B_{x}(x,Q^{2})$ are shown in the
Figs.~\ref{smallQ}~-~\ref{highQ} where they are compared to the predictions
of theoretical models. The results of the interpolation for $x=0.005, 0.01,
0.05 $ and 0.08 are presented in the Fig.\ref{B-xx}.

We would like to comment some "technical" points in our analysis and
results.
\begin{itemize}
\item[{\it i})]
In order to keep its local character to the $x$-slope and to obtain a
maximal possible number of "measurements", we have given only the case
when five points are taken in each elementary bin. We found a weak
dependence of the extracted slopes if the number of points in a bin goes
from $N= 4$ to $N= 6$.
\item[{\it ii})]
In spite of a high accuracy of the recent data from HERA, the dispersion
of the SF-values influences the resulting values of $B_{x}$ and its
uncertainty. For some bins, for example, we were unable to fit
(\ref{localF2})
to the data with $\chi^{2}\leq 1$, so we could not obtain in all cases
reasonable errors. Nevertheless we show these extracted values (only with
$\chi^{2}\leq 3$) in the figures and use them for interpolation to the $x$
under interest because even with those points the extracted set of data
for $B_{x}(x,Q^{2})$ at fixed $x$ is quite poor. Of course, this reduces
the reliability of our results, but only slightly because the total number of
"bad" bins remains small.
\item[{\it iii})]
"Experimental" values of $B_{x}$ at fixed $x$ shown in Fig.~\ref{B-xx} are
obtained by a linear interpolation within the two subsets of local
$x$-slopes, extracted from the HERA and from the fixed target measurements
of SF.
\item[{\it iv})]
One can see in Fig.~\ref{B-xx} that some points deviate strongly from
the groups constituted by the other ones. This is due to the strong
influence of the points (of $F_{2}$ as well as of $B_{x}$) which are at
the ends of the $x$-bins and which also "fall out of a common line"
(see also item {\it ii} above). To solve this problem, a possible way
would be to exclude some of them from the analysis; another way would
be to enlarge the number of points in an elementary bin loosing,
however, the local character of the extracted $x$-slope. A more
detailed analysis of the available data related to more numerous
measurements of the primary observable, $F_{2}(x,Q^{2})$ would be
necessary to obtain more precise data for $B_{x}(x,Q^{2})$.
\item[{\it v})]
In all figures we show for comparison the recently published data (open
circles) on slopes of H1 Collaboration~\cite{BH1-01}. Our
slopes extracted by the method of overlapping bins and those obtained in
\cite{BH1-01} are in a good agreement.
\end{itemize}

%%%%%%%%%%%%%%%%%%%%%%%%%%%%%%%%%%%%%%%%%%%%%%%%%%%%%%%%%%%%%%
%%%%%%%%%%%%%%%%%%%%%%%%%%%%%%%%%%%%%%%%%%%%%%%%%%%%%%%%%%%%%%
%%%%%%%%%%%%%%%%%%%%%%%%%%%%%%%%%%%%%%%%%%%%%%%%%%%%%%%%%%%%%%
\section{Comparison of the slopes data with the models predictions and
conclusions}\label{Comparison}
There are many parameterizations of the proton structure functions which
successfully accommodate for the whole - or a part of - available data
set and in particular for the steep rise of the SF when $x$ decreases for
a large span of $Q^2$ values . Several of them have given explicitly some
hints (see \eg~\cite{Be-95,DLM,TroTyu,DM,CS}) on the existence of a
slowing down of the HERA effect.

Here we concentrate on the three models (SDP, GLP, CSP) considered in the
Section~\ref{Models}
satisfying to Pomeron universality, \ie with Pomeron contribution which is
the same for hadron-hadron and $\gamma$-proton elastic interactions. These
models lead to $\sigma_{tot}^{\gamma p}(s)\propto \ell n^{\mu}(s)$ at
$s\to \infty$ with $\mu=1$ (SDP and GLP models) and with $\mu=2$ as in CSP
model.

All models well describe the extracted data on $B_{x}(x,Q^{2})$,
particularly a growth of this quantity with available $Q^{2}$ when $x$ is
fixed, in spite of the fact that the real Pomeron intercept is constant
and equal one.

\begin{figure}
\begin{center}
\includegraphics*[scale=0.7]{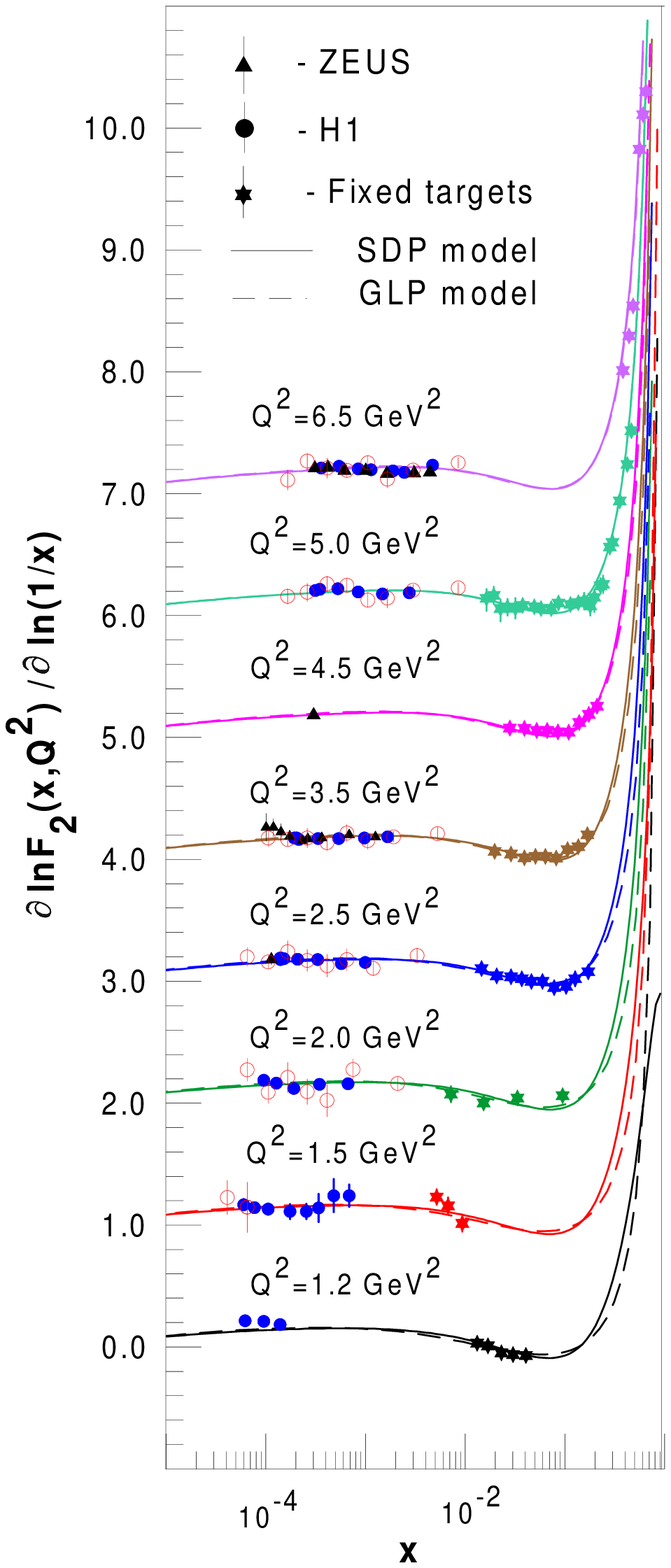}
\caption{$B_x(x,Q^2)$ as function of $x$ at small $Q^{2}$. The open circles
are the data from \cite{BH1-01}. The other notations are given
in the figure} \label{smallQ}
\end{center}
\end{figure}

\begin{figure}
\begin{center}
\includegraphics*[scale=0.7]{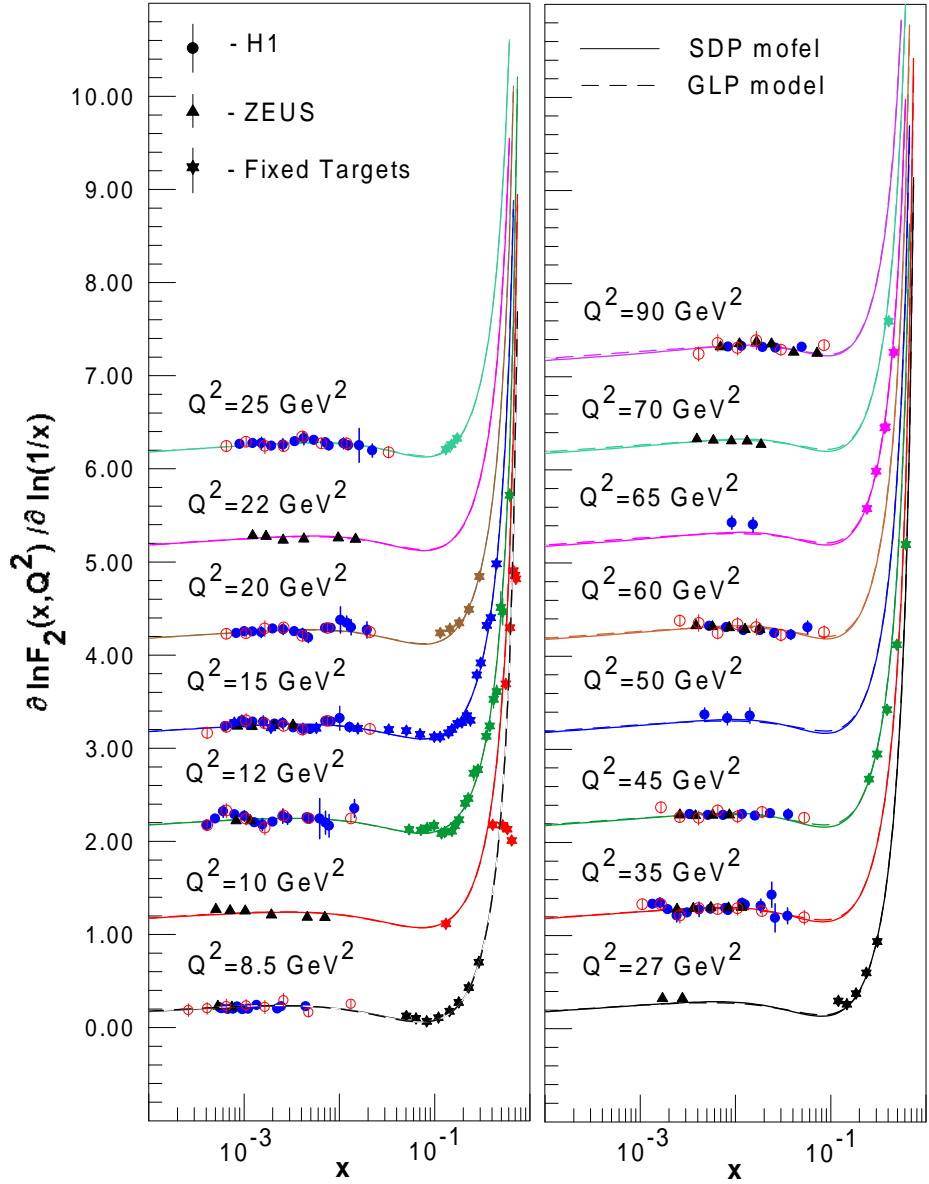}
\caption{The same as in Fig. \ref{smallQ} but for intermediate $Q^{2}$.}
\label{intQ}
\end{center}
\end{figure}

\begin{figure}
\begin{center}
\includegraphics*[scale=0.7]{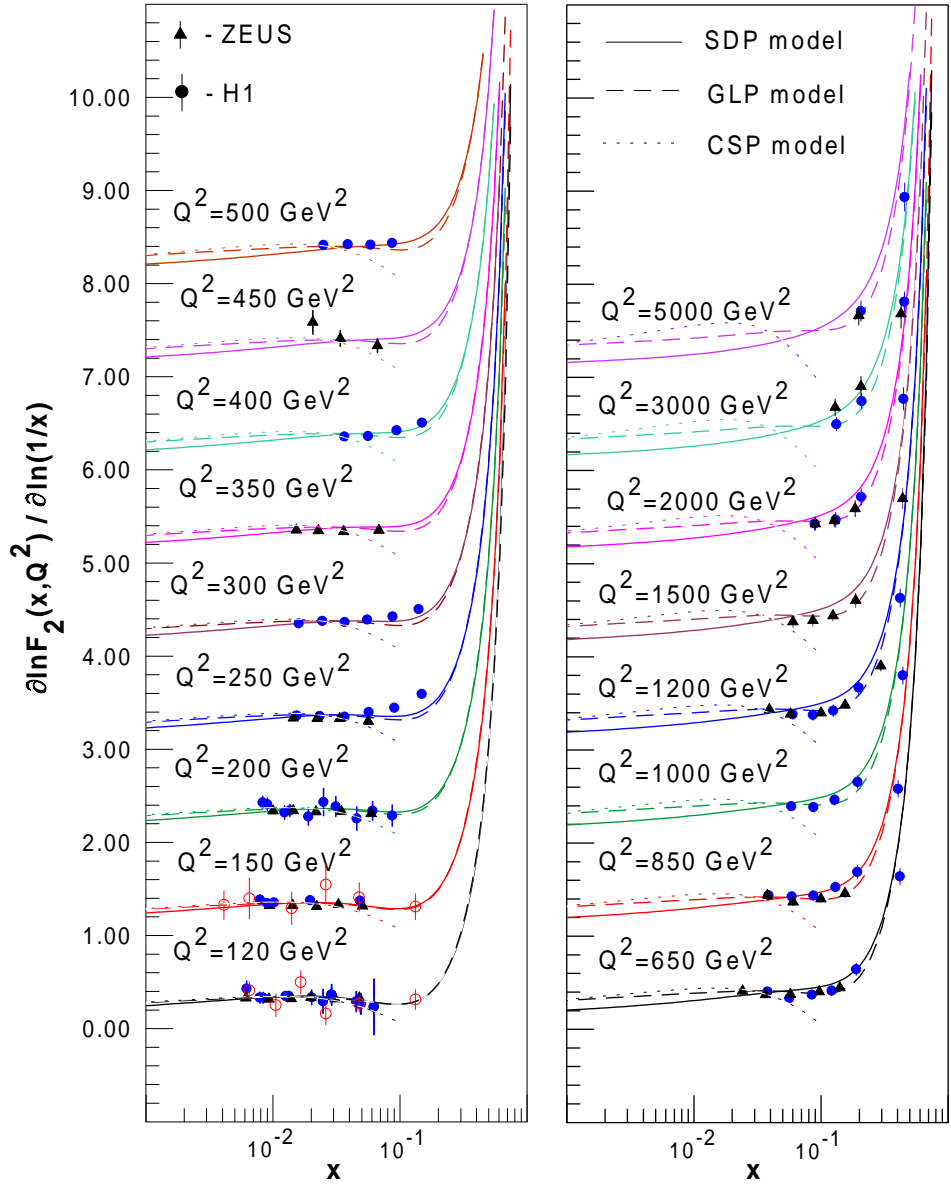}
\caption{The same as in Fig. \ref{smallQ} but for high values of $Q^{2}$.}
\label{highQ}
\end{center}
\end{figure}

\begin{figure}
\begin{center}
\includegraphics*[scale=0.55]{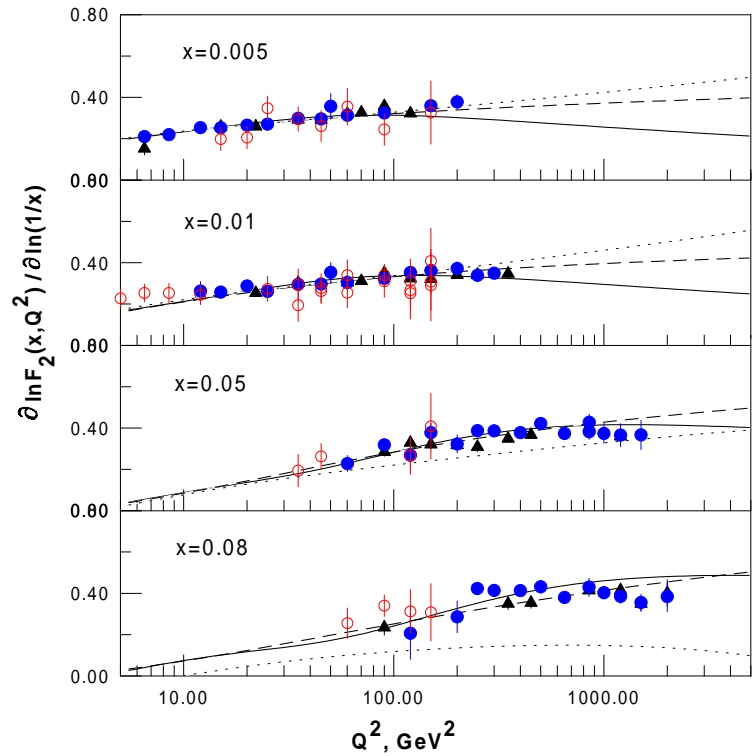}
\caption{$B_x(x,Q^2)$ at fixed $x$ versus $Q^{2}$. The notations
for experimental points and theoretical curves are the same as in
Figs.~\ref{smallQ}-\ref{highQ}. Only data of \cite{BH1-01} closed
to the given values of $x$ are shown.} \label{B-xx}
\end{center}
\end{figure}

One can see in the Figs.~\ref{smallQ}~-~\ref{highQ} and in Fig.~\ref{B-xx}
that the theoretical curves calculated in all models
are closed each to others for small and intermediate $Q^{2}$ while they
are noticeably different for high $Q^{2}\gsim 200$ GeV$^{2}$. As
concerns the CSP model, one must recall that it can be applied in its original
form only for small $x$. That is why the data for $x\gsim 0.05$ are not
described by this model. However, we checked it, it is possible to extend its
ability to describe SF in the large-$x$ region modifying each $i$-term
of~(\ref{Cud-Soy}) by a factor $(1-x)^{B_{i}(Q^{2})}$. Unhappily, an extra
bump-like structure in $B_{x}$ is predicted by such a modified CSP
model which is not supported by available data.

It is seen in Fig.~\ref{B-xx} that the GLP model predicts the growth of
$B_{x}$ at $Q^{2}\to \infty$ to the asymptotic value
$B_{as}=\epsilon\approx 0.454$. A decreasing of $x$-slope predicted in
the SDP and CSP models is observed at such high values of $1/x$ and
$Q^{2}$ that it is problematic to detect the phenomenon in future
experiments. However it seems possible to distinguish SDP and CSP
models from GLP one when the data at fixed $x$ and higher $Q^{2}$ will
be known.

Returning to the often assumed constancy of $B_{x}$ at $x<0.01$, we
would like to note that the 3 considered models exhibit a very weak
$x$-dependence of $B_x$ at small $x$. At the same time $B_{x}(x\to
0,Q^{2})\to 0$ for any fixed $Q^{2}$.

\begin{figure}
\begin{center}
\includegraphics*[scale=0.7]{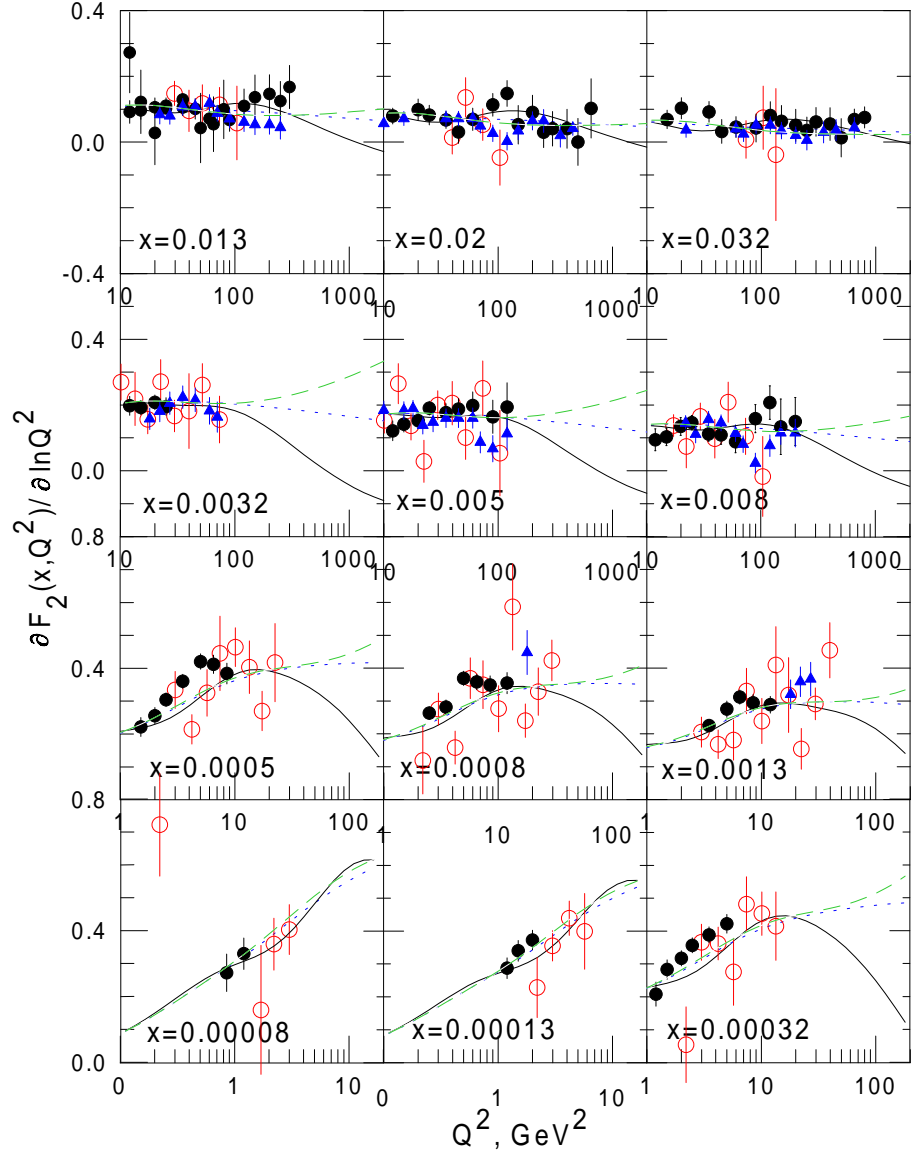}
\caption{Comparison of $Q$-slopes (\ref{B-Q}) at fixed $x$ versus
$Q^{2}$ extracted by the overlapping bins method from ZEUS (full
triangles) and H1 (full circles) $F_{2}$-data with the $Q$-slopes of
\cite{BH1-01} (open circles) and with the predictions of SDP (solid
lines) and GLP (dashed lines) models.} \label{B-Qx}
\end{center}
\end{figure}

\begin{figure}
\begin{center}
\includegraphics*[scale=0.7]{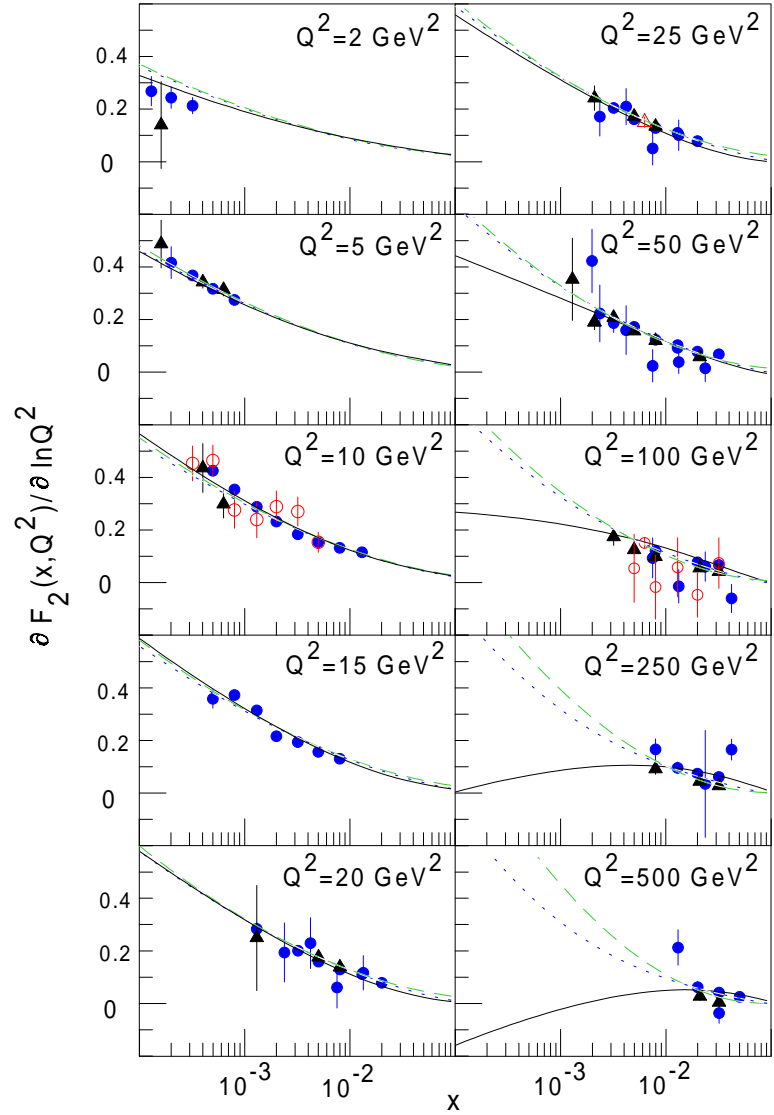}
\caption{Same as in Fig. \ref{B-Qx} but at fixed $Q^2$ versus $x$.}
\label{B-QQ}
\end{center}
\end{figure}

The method of overlapping bins has been applied also in order to
extract from the SF data the other derivative, the $Q-$slope. Again as
for the $x$-slope, we extracted $B_{Q}$ separately from the H1, ZEUS
and Fixed Target data. The results as well as the theoretical curves
calculated in the models under interest are presented in
Fig.~\ref{B-Qx} and Fig.~\ref{B-QQ}. The Caldwell-plot with the
corresponding points and curves for SDP (solid line), GLP (short dotted
line) and CSP (long dotted line) models is presented in Fig.~7. One can
see that all models are in good agreement with the data.
However they predict quite different behaviors outside
the region of available data, in particular at small $x$ and high
$Q^{2}$ (see Fig.~\ref{B-Qx}).

\medskip

Our short {\bf conclusion} is the following. We have shown that the
available data on the proton structure function as well as on the $x$- and
$Q$-slopes are well described in the Regge models that realize the idea of
a universal Pomeron. Pomeron singularity in these models is the same in
hadron-hadron and in photon-hadron elastic forward amplitudes. It
is located in angular momentum plane at $j=1$ not violating the unitarity
restrictions. These models predict a decreasing of $x$-slope and
$\lambda(x,Q^{2})$ for fixed $Q^{2}$ and $x\to 0$.

\begin{figure}
\begin{center}
\includegraphics[scale=0.7]{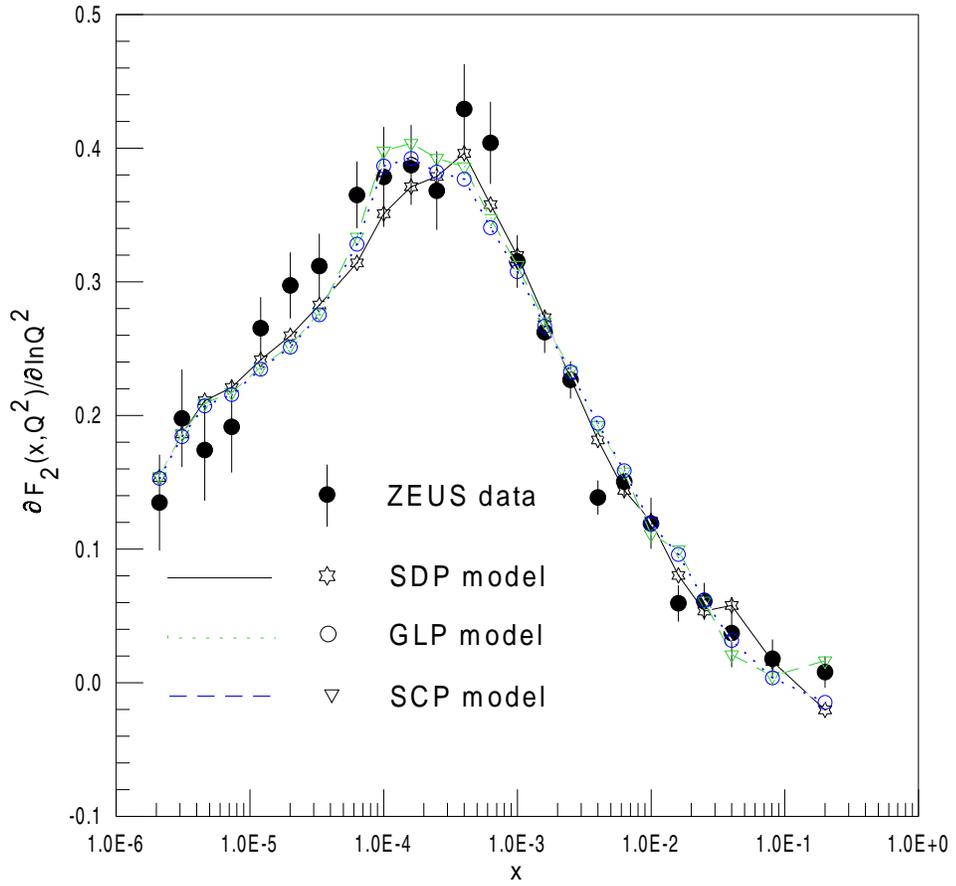}
\caption{The $B_{Q}$-slope at the selected ZEUS points given by the
solid circles (Caldwell plot). $B_{Q}$ calculated in SDP, GLP and CSP
models are presented also.} \label{Caldwell-plot}
\end{center}
\end{figure}

\end{document}